\documentclass[aps,prb,twocolumn,superscriptaddress,showpacs]{revtex4-2}
\usepackage{bm}
\usepackage{amsmath}
\usepackage{amssymb}
\usepackage{graphicx}
\graphicspath{{img_1/}}  
\usepackage{color}

\begin{document}

\title{Interplay of Vibration and Coulomb
Effects in Transport of Spin-Polarized Electrons in a
Single-Molecule Transistor}
\author{A.D. Shkop}
\affiliation{B. Verkin Institute for Low Temperature Physics and
Engineering of the National Academy of Sciences of Ukraine, 47
Prospekt Nauky, Kharkiv 61103, Ukraine}

\author{O.M. Bahrova}
\affiliation{B. Verkin Institute for Low Temperature Physics and
Engineering of the National Academy of Sciences of Ukraine, 47
Prospekt Nauky, Kharkiv 61103, Ukraine}

\author{S.I. Kulinich}
\affiliation{B. Verkin Institute for Low Temperature Physics and
Engineering of the National Academy of Sciences of Ukraine, 47
Prospekt Nauky, Kharkiv 61103, Ukraine}

\author{I.V. Krive}
\affiliation{B. Verkin Institute for Low Temperature Physics and
Engineering of the National Academy of Sciences of Ukraine, 47
Prospekt Nauky, Kharkiv 61103, Ukraine}
\affiliation{Physical Department, V. N. Karazin National University, Kharkiv 61077, Ukraine}
\affiliation{Center for Theoretical Physics of Complex Systems,
Institute for Basic Science (IBS), Daejeon 34051, Republic of Korea}

\begin{abstract}
 Tunnel transport of interacting spin-polarized electrons through a single-level
 vibrating quantum dot in external magnetic field is studied. By using density 
 matrix method, the current-voltage characteristics and the dependence of conductance 
 on temperature of a single-electron transistor were calculated. We found  
 that a lifting of Coulomb blockade in external magnetic field happens in stages. The Franck-Condon steps
 associated with inelastic electron tunneling in our case are doubled  
 due to contribution of two Zeeman-split 
 levels in electron transport. The doubling of steps can be also observed in the presence of
 Coulomb interaction. 
 For strong electron-vibron interaction the temperature dependence of conductance is shown
 to be non-monotonic and anomalous growth of conductance maximum weakly depends both on the
 Coulomb strength and the external magnetic field. 
\end{abstract}

\maketitle

\section*{Introduction}
 Single-electron transistors (SET) and molecular transistors are of great interest nowadays. 
 In particular single-molecule transistors can be useful for molecular spectroscopy.
 A specific feature of molecular transistors is the influence of vibration degrees of 
 freedom on electron transport. 
 Effects of electron-vibron
 interaction in electron transport through metal-molecule-metal junctions had
 been already observed in several experiments \cite{Park,Pasupathy,Kubatkin} 
 (see also recent years
 publications \cite{Monthioux,Burzuri,Martinez-Blanco}).

There are two different regimes of electron transport in molecular transistors -- 
the shuttling regime and vibronic regime. 
The electron shuttling \cite{Shuttle}  takes place for weak electromechanical coupling and when the 
interaction of vibrons 
with the environment is not sufficiently strong to prevent mechanical oscillations of molecule's center of mass. 
For strong 
interaction of vibrons with the environment strong dissipation in mechanical subsystem appears. In this
case oscillations are damped and vibration effects are associated with vibron-assisted electron tunneling 
(vibronic regime) \cite{Glazman,Mitra,Flensberg}. 

We consider a single-molecule transistor in the latter of these two regimes. It was shown earlier 
(see e.~g. Refs. \cite{Koch,Galperin}) that the energy of an electron in vibrating quantum dot is decreased 
(polaronic shift) due to electron-vibron interaction. The vibrons allow the electron to tunnel inelastically through 
the junction and the current-voltage characteristics of molecular transistor at low temperatures are described
by step-like functions with the so-called Franck-Condon (FC) steps. Each step of current appears when 
a new inelastic 
channel enters the conducting $"$window$"$ when the bias voltage is increased. For large electron-vibron coupling 
constants 
the tunnel current at low temperatures and bias voltages is strongly suppressed (Franck-Condon blockade \cite{Koch}). 

The influence of strong electron-vibron interaction on electron transport is manifested in the dependence of 
conductance  
on bias voltage (Franck-Condon blockade of current at low bias voltages) or in the temperature dependence
of conductance.  For sequential electron tunneling the dependence of linear conductance on temperature 
is not monotonic~\cite{Krive} 
and at low temperatures it is additionally suppressed  (due to polaronic narrowing of the dot level width) compared to 
the conductance of SET with immovable quantum dot.

Transport of spin-unpolarized electrons in single-molecule transistors is theoretically well
studied phenomenon. 
It is interesting to generalize the model, which describes vibration effects
in transport of spin-unpolarized electrons, to the case when electrons are
spin-polarized (see also~\cite{Bruggemann}). One can control the transport of spin-polarized electrons by an
external magnetic field, thus corresponding device can be useful in spintronics. 
In our model (see Fig.~1a) spintronic SET is represented by a single-level 
vibrating quantum dot (QD) tunnel coupled to magnetic electrodes. The magnetization vectors of 
the leads are set to be anti-parallel.
Then to get non-zero current for fully spin-polarized leads one needs to introduce 
external magnetic field. External magnetic field in our device is perpendicular to 
the plane of magnetization vectors. 
The system exhibits a "spin blockade"~\cite{Shekhter} property  -- for zero external magnetic 
field the electric current is blocked. 
Magnetic field induces electronic spin-flips in the dot and lifts spin blockade. 
Coulomb correlations are
known to be important in electron transport in single-molecule transistors and in our theory
we also take into account 
electron-electron interaction in the dot. 
The presence and interplay of 
magnetic properties  and Coulomb interaction could strongly 
affect step-like current-voltage characteristics of our spintronic device.
\begin{figure}\centering 
\includegraphics[scale=0.5]{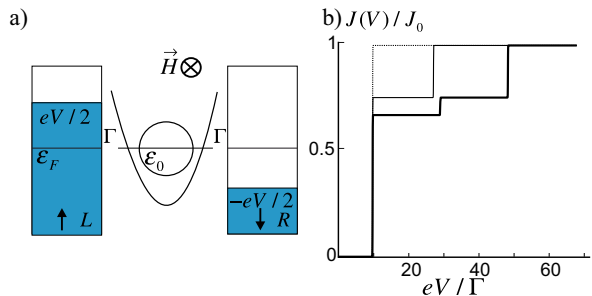}
\caption{a) A single-molecule transistor modelled by a vibrating quantum dot (QD)
(with energy level $\varepsilon_0$) placed between two spin-polarized electrodes with the chemical potentials $\mu_{L,R}=\pm eV/2$ counted from Fermi level. A schematic picture of QD with Zeeman-split energy level coupled to the leads ($\Gamma$ is the level width) is presented. The magnetization vectors of electrodes are anti-parallel 
(depicted as arrows) and electrons in the leads are fully spin-polarized. An external magnetic field perpendicular 
to the plain of magnetization is applied. b) The current-voltage dependences in the absence of electron-vibron 
interaction, $\lambda=0$, for different values of  Coulomb coupling constant: $U=0$ 
(dotted), U= $9\Gamma$ (solid thin), U=$20\Gamma$ (solid thick) and magnetic field ($h=g \mu_B H$) is $h=10\Gamma$. 
The plots show that Coulomb blockade of current is lifted in stages. 
For solid thick curve ($U=20\Gamma$) the jumps of current 
associated with opening of tunneling channels are clearly seen. For the solid thin curve ($U=9\Gamma$) the number of steps of the current is reduced due to the condition $U \leqslant h$. All graphs are calculated for small 
temperature $T=0.01 \Gamma$, the current is shown in units $J_0=e\Gamma/2$.}
\end{figure}

All calculations in the present paper are done in a weak tunneling limit, 
which is relevant for description of electron transport in 
experiments with single-molecule transistors. We use the approximation of sequential electron tunneling,
i.e. phase coherence for electron tunneling processes is absent. However, spin dynamics
in our device is fully coherent and we will use density matrix method to calculate the average current. 
For sequential electron tunneling total density matrix can be factorized to the density matrix of QD and 
equilibrium density
matrices of the leads. This allows one to solve Liouville-von Neumann equation
by perturbation over level width $\Gamma\ll \text{max}\{T,eV\}$, where $T$ is the temperature of the electrodes ($k_B=1$), $e$ is the elementary charge, $V$ is the bias voltage.

The current-voltage characteristics for different model parameters and the  dependence of conductance on 
temperature, external magnetic field and the strength of Coulomb interaction are calculated numerically. 
We observe the lifting of Coulomb blockade in our system in stages. On top of the steps of 
Coulomb blockade lifting  we predict the doubling and suppression of FC steps in the presence of electron-vibron interaction.  

The temperature dependence of maximal conductance for 
strong electron-vibron interaction is non-monotonic at intermediate temperatures analogous to the results of 
the work~\cite{Monthioux}, obtained for spinless electrons. This anomalous behaviour is also present for non-zero
Coulomb interaction. 
The growth of conductance with the increase of temperature is maximal when external magnetic field and Coulomb interaction are of the order of vibron energy $\hbar\omega$ ($\omega$ is the frequency of dot vibrations).

\section{Hamiltonian and kinetic equation for density matrix}

The Hamiltonian of our system consists of four terms
\begin{equation}\label{1}
\mathcal{H}= H_l+ H_{d}+H_{v-d}+ H_{tun},
\end{equation}
where $H_l$ is the Hamiltonian of non-interacting electrons in the
leads
\begin{equation}\label{2}
H_l=\sum_{k,\alpha}\varepsilon_{k,\alpha}
c^\dag_{k,\alpha}c_{k,\alpha},
\end{equation}
where $c_{k,\alpha}^\dag (c_{k,\alpha})$ is the creation (annihilation)
operator (with the standard anti-commutation
relations) of an electron with the momentum $k$
and the energy $\varepsilon_{k,\alpha}$ in the electrode $\alpha=L,R$ (due to full spin-polarization the lead index corresponds to the spin index $\uparrow,\;\downarrow$). $H_{d}$ is the QD Hamiltonian
\begin{equation}\label{3}
H_{d}=\sum_\sigma \varepsilon_0 d_\sigma^\dag d_\sigma-\frac{g\mu_B
H}{2}(d_\uparrow^\dag d_\downarrow+d_\downarrow^\dag
d_\uparrow)\\
+U d_\uparrow^\dag d_\uparrow d_\downarrow^\dag
d_\downarrow.
\end{equation}
Here $d_\sigma^\dag(d_\sigma)$ is the creation (annihilation)
operator of electron state on the dot with the energy
$\varepsilon_0$ and spin projection $\sigma=\uparrow,\downarrow$,
$U$ is the strength of electron-electron interaction in the dot,
$H$ is the external magnetic field,  $g$
is the gyromagnetic ratio, $\mu_B$ is the Bohr magneton.

Hamiltonian $H_{v-d}$ describes the vibronic subsystem and electron-vibron interaction. It reads
\begin{equation}\label{4}
H_{v-d}=\hbar\omega b^{\dag}b+\varepsilon_{int}
(b^{\dag}+b)(d_\uparrow^\dag d_\uparrow+d_\downarrow^\dag d_\downarrow),
\end{equation}
where $b^{\dag} (b)$ creates (annihilates) a vibron in the dot, $\omega$ is the angular frequency of dot vibrations, $\varepsilon_{int}$ is the characteristic interaction energy. 

In the case of full spin polarization of the leads the Hamiltonian
$H_{tun}$ describing the tunnelling of electrons  between the
dot and the leads can be represented in the form
\begin{equation}\label{5}
H_{tun}=t_L\sum _k c^\dag_{k,L}d_\uparrow+t_R\sum _k
c^\dag_{k,R}d_\downarrow+\text{H.c.},
\end{equation}
where $t_\alpha$ is the tunnelling amplitude. In what follows we
restrict ourselves to the case of symmetric junction $t_L=t_R=t_0$.

It is convenient to perform two transformations with the
Hamiltonian $\mathcal{H}$: (i) the canonical transformation
of the dot electronic operators $d_\sigma$
\begin{equation} \label{6}
d_{\uparrow} =\frac{1}{\sqrt{2}}(a_1+a_2),\,d_{\downarrow} =
\frac{1}{\sqrt{2}}(-a_1+a_2),
\end{equation}
and (ii) the unitary transformation of the Hamiltonian,
$\Lambda \mathcal{H} \Lambda^\dag$, with $\Lambda=\exp[\lambda (b^{\dag}-b)(n_1+n_2)]$, where
$n_s=a_s^\dag a_s$, $s=1,2$, with $\lambda$ playing the role of dimensionless electron-vibron coupling.
If we set $\lambda=\varepsilon_{int}/ \hbar\omega$ these transformations lead to the following diagonal 
form of the Hamiltonian $H_{d}$ and the Hamiltonian $H_{v-d}$
\begin{eqnarray}\label{U_lambda}
&&H_{d}\rightarrow H_{d}=\sum_{s=1}^2 \varepsilon_s a_s^\dag
a_s+(U-2\lambda^2\hbar\omega)n_1n_2,\\
&&H_{v-d}\rightarrow H_v=\hbar\omega b^{\dag}b,\label{Hvibr}
\end{eqnarray}
where $\varepsilon_{1,2}=\varepsilon_0-\lambda^2\hbar\omega \pm
h/2$ and $h=g\mu_B H$ is the external magnetic field in energy units. Dot energy level and the Coulomb interaction strength are now 
renormalized by electron-vibron interaction. The case when $U-2\lambda^2\hbar\omega<0$ needs 
special considerations and in what follows we will assume positive renormalized Coulomb interaction. 
The vibration-induced shift of energy level is known as the polaronic shift (see e.~g.~\cite{Krive}).
Since the level energy can always be changed by gate voltage and tuned to the Fermi energy
to get maximum current, in what follows we use the
notation
$\varepsilon_0$ instead of $\varepsilon_0-\lambda^2\hbar\omega $ and $U$ instead 
of $U-2\lambda^2 \hbar\omega$. After transformations (i) and (ii) the tunnel
Hamiltonian, $H_{tun}$, (Eq.~(\ref{5})), reads
\begin{equation} \label{8}
H_{tun}=\frac{t_0}{\sqrt{2}}\sum_{k,\alpha}\text{e}^{-\lambda (b^{\dag}-b)}
c^\dag_{k,\alpha}(j_\alpha a_1+a_2)+\text{H.c.},
\end{equation}
where $j_{L,R}=\pm 1$.

The Liouville-von Neumann equation for the density matrix (here and below $\hbar=1$)

\begin{equation}\label{10}
\frac{\partial \rho}{\partial t}+i[H_0+H_{tun},\rho]=0,
\end{equation}
with the unperturbed Hamiltonian $H_0=H_l+H_v+H_{d}$, has the formal solution
\begin{equation}\label{11}
\rho(t)=\rho_{-\infty}-i\int\limits_{-\infty}^t dt'\text{e}^{-i
H_0(t-t')}[H_{tun},\rho(t')]\text{e}^{i H_0(t-t')},
\end{equation}
where $\rho_{-\infty}=\rho(t\rightarrow-\infty)$.
The quantum consideration of electron subsystem in our model is based on the
assumption that the density matrix of the system can be factorized
to a direct product of the leads density matrix, vibron  density
matrix and density matrix of the dot
\begin{equation}\label{9}
\rho\approx\rho_l\otimes\rho_v\otimes\rho_{d},
\end{equation}
and we assume  that the density matrix of the leads and vibron
density matrix are equilibrium. This approximation corresponds to sequential electron 
tunnelling regime and strong coupling of vibrons to the heat bath. Our
approach
is valid for $\Gamma \ll \text{max}\{T,eV\} $. Note that the approximation, Eq.~(\ref{9}), does not allow one
to consider the resonance phenomena (such as the Kondo effect,
co-tunnelling etc.), but it is an adequate method for treating stochastic
electron tunneling and coherent spin dynamics.

Substituting Eqs.~(\ref{11}),~(\ref{9}) into
Eq.~(\ref{10}) and using symbol "Tr" to denote the tracing over both the electronic degrees of
freedom in the leads and vibronic degrees of freedom in the dot
one gets
\begin{equation}\label{12}
\begin{array}{c}
\dfrac{\partial \rho_{d}}{\partial t}+i[H_{d},\rho_{d}]=\\ 
-\text{Tr}\int\limits_{-\infty}^t dt'[H_{tun}, \text{e}^{-i
H_0(t-t')}[H_{tun},\rho(t')]\text{e}^{i H_0(t-t')}].
\end{array}
\end{equation}
Within our assumption, Eq.~(\ref{9}), the average with the unperturbed Hamiltonian 
of the electrodes yields
\begin{equation}\label{13}
\langle c^\dag_{k,\alpha}c_{k',\alpha'}\rangle=
f_\alpha(\varepsilon_{k,\alpha})
\delta_{k,k'}\delta_{\alpha,\alpha'},
\end{equation}
where $f_\alpha(\varepsilon)=(\exp((\varepsilon-\mu_\alpha)/T)+1)^{-1}$
is the Fermi-Dirac distribution function,  $\varepsilon$ is the energy and $\mu_\alpha= j_\alpha
(eV/2)$ is the electrochemical potential in the lead $\alpha$ counted from $\varepsilon_F$. 

The average of the product of exponential functions $\exp(\pm\lambda [b^{\dag}(t)-b(t)])$ taken
at different times $t,t^{\prime}$ with the unperturbed Hamiltonian of vibronic subsystem, Eq.~(\ref{Hvibr}),
within the assumption that vibrons are in equilibrium with the heat bath at
temperature $T$, takes the form (see e.~g.~\cite{Mahan,Flensberg})
\begin{equation}\label{14}
\langle \text{e}^{\mp\lambda [b^{\dag}(t)-b(t)]}\text{e}^{\pm\lambda
[b^{\dag}(t^\prime)-b(t^\prime)]}\rangle_0=\sum\limits_{n=-\infty}^{+\infty} A_n \text{e}^{-i\omega n(t-t^\prime)},
\end{equation}
where
\begin{equation}
A_n=\text{e}^{-\lambda^2(1+2n_B)}I_n(z)\text{e}^{n\omega/2T}.
\end{equation}
Here $I_n(z)$ is the
modified Bessel function of the first kind, $z=2\lambda^2\sqrt{n_B(1+n_B)}$, $n_B$ is the 
Bose-Einstein distribution function 
$n_B=(\exp(\omega/T)-1)^{-1}$, $\omega$ is the quantum of mechanical oscillations. We suppose that the temperatures of
electron subsystem in the leads and vibron subsystem in the dot
are the same.

With the help of Eqs.~(\ref{13}),~(\ref{14}), Eq.~(\ref{12}) is transformed to
\begin{eqnarray}\label{15}
\dfrac{\partial \rho_{d}}{\partial
t}+i[H_{d},\rho_{d}]=\dfrac{\Gamma}{4}\sum\limits_{n,\alpha}A_n\int
d\varepsilon
\frac{1}{2\pi}\int d\tau\times \nonumber \\
\left\{\text{e}^{i \varepsilon^+_n \tau}[1-f_\alpha(\varepsilon)]a_{\alpha}\text{e}^{-i
H_{d}\tau} \rho_{d}(t-\tau)a_{\alpha}^{\dag}\text{e}^{i H_{d}\tau} \right.\nonumber\\
+\text{e}^{-i\varepsilon^-_n \tau}f_\alpha(\varepsilon)a_{\alpha}^{\dag}\text{e}^{-i
H_{d}\tau} \rho_{d}(t-\tau)a_{\alpha}\text{e}^{i
H_{d}\tau}\nonumber\\
-\text{e}^{-i\varepsilon^+_n\tau}[1-f_\alpha(\varepsilon)]a_{\alpha}^{\dag}\text{e}^{-i H_{d}\tau}
a_{\alpha}\rho_{d}(t-\tau)\text{e}^{i
H_{d}\tau}\nonumber\\
-\left.\text{e}^{i \varepsilon^-_n\tau}f_\alpha(\varepsilon)a_{\alpha}\text{e}^{-i
H_{d}\tau} a_{\alpha}^{\dag}\rho_{d}(t-\tau)\text{e}^{i
H_{d}\tau}\right\} +\text{H.c.},
\end{eqnarray}
where we denote $\varepsilon^{\pm}_n=\varepsilon \pm n \omega $
and $a_{\alpha}=j_{\alpha}a_1+a_2$.
Here $\Gamma=2\pi\nu t_0^2$ is the level width of electron state
in the dot, $\nu$ is the density of states in the leads, which we
assume to be energy independent (wide-band approximation).

The density operator $\rho_{d}$ acts in the Fock space of a single-electron level. 
The matrix elements of density operator are
\begin{eqnarray}\label{16}
\rho_0=\langle 0\vert\rho_{d}\vert 0\rangle,\rho_s=\langle
s\vert\rho_{d}\vert s\rangle, s=1,2,\nonumber\\
\rho_D=\langle D\vert\rho_{d}\vert D\rangle,\;\; \rho_{ss^\prime}=\langle
s\vert\rho_{d}\vert s^\prime\rangle, s\neq s^\prime=1,2,
\end{eqnarray}
where $\vert 0\rangle$ is
the vacuum state, $\vert s\rangle=a_s^\dag \vert 0\rangle$, and the state $\vert D
\rangle=a_1^\dag a_2^\dag \vert 0\rangle$ corresponds to doubly occupied
dot.

In a steady-state regime the reduced density operator $\rho_{d}$
is time independent. In this case 
one gets from Eqs.~(\ref{15}),~(\ref{16}) the following system of equations for the matrix elements
\begin{eqnarray}
&&-\rho_0\left[D_+(\varepsilon_1)+D_+(\varepsilon_2)\right]+
\rho_1 [1-C_+(\varepsilon_1)]\nonumber\\
&&+\rho_2[1-C_+(\varepsilon_2)]- \rho_+
[C_-(\varepsilon_1)+C_-(\varepsilon_2)]=0,\label{17}\\
&&\rho_0 D_+(\varepsilon_1)-\rho_1 F_1+\rho_D
[1-C_+(\varepsilon_2+U)]\nonumber\\
&&+\rho_+[C_-(\varepsilon_1)+D_-(\varepsilon_2+U)]=0,\\
&&\rho_0 D_+(\varepsilon_2)-\rho_2 F_2+\rho_D
[1-C_+(\varepsilon_1+U)]\nonumber\\
&&+\rho_+[C_-(\varepsilon_2)+D_-(\varepsilon_1+U)]=0,\\
&&\rho_1 D_+(\varepsilon_2+U)+\rho_2
D_+(\varepsilon_1+U) \nonumber\\
&&-\rho_D[1-C_+(\varepsilon_1+U)]-\rho_D [1-C_+(\varepsilon_2+U)]\nonumber\\
&&-\rho_+[D_-(\varepsilon_1+U)+D_-(\varepsilon_2+U)]=0,\\
&&h\rho_+=\Gamma \rho_-F,\\
&&h\rho_-=-\Gamma(\rho_+F-L/2).
\label{18}
\end{eqnarray}
In Eqs.~(\ref{17})-(\ref{18}) we use the following notations
\begin{eqnarray}
\rho_+=(\rho_{12}+\rho_{21})/2,\\
\rho_-=i(\rho_{12}-\rho_{21})/2,\\
C_\pm (\varepsilon)=\dfrac{1}{2}\sum_n A_n [f_L(\varepsilon-n\omega)
\pm f_R(\varepsilon-n\omega)],\label{19a}\\
D_\pm (\varepsilon)=\dfrac{1}{2}\sum_n A_n [f_L(\varepsilon+n\omega)
\pm f_R(\varepsilon+n\omega)],\\
L=\rho_0[D_-(\varepsilon_1)+D_-(\varepsilon_2)]
\nonumber\\
+\rho_1[C_-(\varepsilon_1)+
D_-(\varepsilon_2+U)]+\nonumber \\
+\rho_2[C_-(\varepsilon_2)+D_-(\varepsilon_1+U)]
\nonumber\\
+\rho_D[C_-(\varepsilon_1+U)+C_-(\varepsilon_2+U)],\\
F_{1,2}=1-C_+(\varepsilon_{1,2})+D_+(\varepsilon_{2,1}+U),
\\
F=(F_1+F_2)/2.\label{19b}
\end{eqnarray}
Notice, that the system, Eqs.~(\ref{17})-(\ref{18}), satisfies (as it should be)
the normalization condition, $\rho_0+\rho_1+\rho_2+\rho_D=1$.

\section{Current of spin-polarized
electrons in molecular transistor}

The "left/right" current in the system is defined by a standard
formula
\begin{equation}\label{20}
J_\alpha=- e \text{Tr}\left(\rho\frac{\partial N_\alpha}{\partial
t}\right), 
\end{equation}
where $N_\alpha=\sum_{k}c_{k,\alpha}^\dag c_{k,\alpha}$ and $e$ is the elementary charge.
The trace is taken over all degrees of freedom of the system. With the help of
Eqs.~(\ref{8}),~(\ref{11}) the current can
be presented in the form
\begin{equation}\label{21}
J_\alpha=\text{Tr}\int\limits_{-\infty}^tdt'\text{e}^{i
H_0(t-t')}I_\alpha\text{e}^{-i
H_0(t-t')}[H_{tun},\rho]
+\text{c.c.},
\end{equation}
where
\begin{equation}
I_\alpha=\frac{et_0}{\sqrt{2}}\text{e}^{-\lambda(b^{\dag}-b)}( j_\alpha a_1+
a_2)\sum_k c_{k,\alpha}^\dag.
\end{equation}
It is convenient to define the total current as the half sum
of left and right currents
$J=(J_L+J_R)/2$. Then the direct calculations lead to the
following relation
\begin{equation}\label{22}
J=e\Gamma(\rho_+F-L/2)=-eh\rho_-.
\end{equation}
As a result the expression for the current takes the form
\begin{equation}\label{Current}
\frac{J}{J_0}=\frac{h^2}{h^2+\Gamma^2F^2}L,
\end{equation}
where $J_0= e\Gamma/2$. Note that from
Eqs.~(\ref{19a})-(\ref{19b}) and Eq.~(\ref{Current}) it is evident that $J(V=0)=0$ and $J(h=0)=0$, as it should be. The maximal current at $V\rightarrow \infty$ for a finite $U$ reads $J_m=J_0 h^2/(h^2+\Gamma^2)$ 
and it tends to $J_0$ only when $h \gg \Gamma$.

The system of Eqs.~(\ref{17})-(\ref{18}) forms a complete
set of equations for calculating the electron  current,
Eq.~(\ref{Current}). It can be simplified in certain limiting cases. In the
numerical and analytical analysis we restrict ourselves to the
case, when the value of the dot level energy 
coincides with the Fermi energy in the leads, $\varepsilon_0=0$. This condition imposes a certain symmetry
on the coefficients $C_\pm(\varepsilon_s),
D_\pm(\varepsilon_s)$
\begin{equation}\label{24}
C_-(\varepsilon_{1,2})=D_-(\varepsilon_{2,1}),\;
C_+(\varepsilon_{1,2})+D_+(\varepsilon_{2,1})=1.
\end{equation}
These properties allow one to obtain relatively simple expressions
for the current in the limiting cases of negligibly small, $U\rightarrow 0$, 
and strong, $U\rightarrow \infty$, 
electron-electron interaction.

If we neglect the Coulomb interaction of electrons in the dot,
$U=0$, the expression for current takes the form
\begin{equation}\label{25}
\frac{J}{J_0}=\frac{h^2\Omega_1}{h^2+\Gamma^2\Omega_2\Omega_3},
\end{equation}
where
\begin{eqnarray}\label{26}
\Omega_1=\frac{2\left[D_+(\varepsilon_1)D_-(\varepsilon_2)+
D_+(\varepsilon_2)D_-(\varepsilon_1)\right]}{D_+(\varepsilon_1)+
D_+(\varepsilon_2)},\\
\Omega_{2,3}=D_+(\varepsilon_1)\pm D_-(\varepsilon_1)+D_+
(\varepsilon_2)\mp D_-(\varepsilon_2).
\end{eqnarray}

Consider now the case of low temperatures, $T\rightarrow 0$. 
The vibron excitations
are suppressed ($n_B\rightarrow0$). At zero temperature the total current equals zero at $eV<h$.
Only when the bias voltage is large enough the
electron can tunnel through the dot with the
emission of vibrons. In this limit the Eq.~(\ref{25}) yields the expression for total current for 
$eV\geqslant h$ in the form as follows
\begin{equation}\label{K}
\frac{J}{J_0}\simeq \frac{2K_+ K_-}{K_++K_-}
\left(\frac{h^2}{h^2+\Gamma^2 K_+K_-}\right),
\end{equation}
where
\begin{equation}\label{Kplusminus}
K_{\pm}=K(N_{\pm})=\text{e}^{-\lambda^2}\sum_{n=0}^{N_{\pm}}
\frac{\lambda^{2n}}{n!},
\end{equation}
and
\begin{equation}\label{Npm}
N_{\pm}=\left[\frac{eV\pm h}{2\omega}\right].
\end{equation}
Notation $[...]$ defines the integer part. Non-analytic dependence of the function $K_{\pm}$
on bias voltage leads to the set of current steps associated with opening of 
inelastic channels formed by two Zeeman-split energy levels. It is seen from 
Eqs.~(\ref{K})-(\ref{Npm}) that the number of FC steps per interval $2\omega$ (we consider symmetric bias voltage) in the current-voltage characteristic for our system is doubled. The doubling of the number of FC steps either in the limit $U=0$ or for non-zero $U$ is confirmed by the numerical simulations (see next section for discussion). However this is not the case for strong fields $h \gg \omega$. For $h \gg  \omega$ one gets $K_+ \gg K_-$, $K_+\rightarrow 1$, 
\begin{equation}\label{Kh}
\dfrac{J}{J_0}\simeq 2K_-\left(\frac{h^2}{h^2+\Gamma^2 K_-}\right).
\end{equation}
Therefore, the appearing of the steps is governed only by the jumps of the function $K_-$ and the doubling of steps is not observed. Nevertheless, the steps have doubled height.

Now we proceed to another limiting case -- Coulomb blockade regime. When $U\rightarrow\infty$ the double
electron occupation of the dot level is forbidden, $\rho_D=0$. In this
case the expression for current takes the form
\begin{equation}\label{28}
\frac{J}{J_0}=\frac{h^2\Omega_4}{h^2\Omega_5+(\Gamma/2)^2\Omega_6},
\end{equation}
where
\begin{eqnarray}\label{29}
\Omega_4=D_+(\varepsilon_1)D_-(\varepsilon_2)+
D_+(\varepsilon_2)D_-(\varepsilon_1),\\
\Omega_5=\frac{[D_+(\varepsilon_1)+D_+(\varepsilon_2)]^2-D_+(\varepsilon_1)
D_+(\varepsilon_2)}{D_+(\varepsilon_1)+D_+(\varepsilon_2)},\\
\Omega_6=[D_+(\varepsilon_1)+D_+(\varepsilon_2)]G_1G_2\nonumber\\
-D_+(\varepsilon_1)[D_+^2(\varepsilon_2)-D_-^2(\varepsilon_2)]\nonumber\\
-D_+(\varepsilon_2)[D_+^2(\varepsilon_1)-D_-^2(\varepsilon_1)].
\end{eqnarray}

At $eV\geqslant h$ the 
behaviour of the current at low temperatures $T\rightarrow 0$ is determined by the expression
\begin{equation}\label{K_infty}
\frac{J}{J_0}\simeq\frac{K_+K_-(K_++K_-)}{(K_++K_-)^2-K_+K_-}
\left[\frac{h^2}{h^2+(\Gamma/2)^2K}\right],
\end{equation}
where
\begin{equation}\label{31}
K=\frac{K_+K_-(K_++K_-)^2}{(K_++K_-)^2-K_+K_-}.
\end{equation}
The doubling of FC steps is also present in the limit $U\rightarrow \infty$. However, for $h \gg \omega$ one gets (compare with Eq.~(\ref{Kh}))
\begin{equation}\label{Kh_infty}
\dfrac{J}{J_0}\simeq K_-\left[\frac{h^2}{h^2+(\Gamma/2)^2K_-}\right],
\end{equation}
and the doubling of the number of steps does not occur in this limit (see also the case of non-interacting electrons for $h \gg \omega$).

We notice also that finite temperature results in appearance of
 current at $eV<h$ due to the inelastic processes of vibron absorption.

\section{Numerical results and discussion}

In Figs.~1b,~2a,b,~3a,b we plotted current-voltage characteristics of our spintronic transistor. 
At first we discuss the transport properties of our device when electron-vibron interaction
is absent ($\lambda=0$). The current-voltage dependences $J(V)$ for different values of Coulomb interaction 
constant (for $U=0$ and for two finite values of interaction strength) 
are shown in Fig.~1b. 
Zeeman splitting of level energy in external magnetic 
field provides two channels for electron tunneling -- the tunneling through the state with the upper, $\varepsilon_1$, and with the lower, $\varepsilon_2$, energy. For non-zero Coulomb interaction electronic correlations suppress the probability 
of elastic tunneling processes through these channels. The probability of tunneling increases, when the Coulomb blockade for electron tunneling is lifted. When the double occupation of the dot level is possible, two additional channels for electron tunneling appear.
These channels correspond to electron tunnelling through doubly occupied state with the excitation energy $\varepsilon_{2,1}+U$ when the levels $\varepsilon_{2,1}$ are occupied. The energy of the doubly occupied electron level is  $\varepsilon_{\uparrow \downarrow}=\varepsilon_1+\varepsilon_2+U=2\varepsilon_0+U=U$, and does not depend on external magnetic field (as it should be). The opening of the channels for electron tunneling at the corresponding threshold bias voltages 
at low temperatures leads to the jumps of current, clearly seen in $J(V)$ plots (Fig.~1b).
Although the Coulomb blockade can be lifted by the increasing of bias voltage, 
this process occurs in stages, 
demonstrating a new property of our system. This type of Coulomb blockade lifting is shown
in the current-voltage dependence for $U>h$ 
(solid thick curve in Fig.~1b). In our symmetric geometry of Zeeman-split levels, $\varepsilon_0=0$, the first stage ends at the threshold voltage $eV=2(U- h/2)$, and the second one-- at $eV=2(U+ h/2)$. For the case $U\leqslant h$ the number of steps of the current is reduced (solid thin curve in Fig.~1b). 
Indeed, now the double occupation of the dot when the lower energy level 
($\varepsilon_2$) is occupied is already possible at the threshold voltage for the current and the Coulomb blockade lifting happens in a single stage.

When electron-vibron interaction is taken into account (see Fig.~2a), the effect of  
Coulomb blockade lifting is also present. 
The only difference is that now Coulomb interaction constant is renormalized by 
electron-vibron coupling (see Eq.~(\ref {U_lambda})).
Furthermore, the current-voltage characteristics demonstrate additional (Franck-Condon) steps.

\begin{figure}\centering 
\includegraphics[scale=0.5]{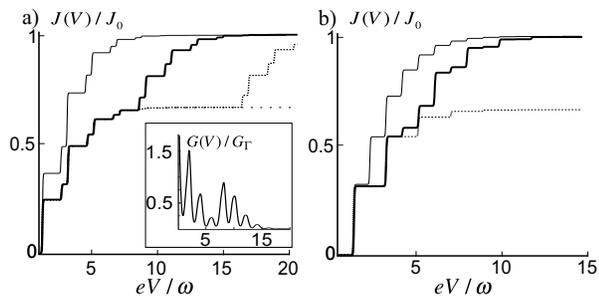}
\caption{a) The current-voltage dependences, when electron-vibron interaction is taken into account, $\lambda=1$.
The external magnetic field is small, $h= 0.25\omega$, ($\omega$ is oscillation
energy quantum). 
The curves are calculated for different values of Coulomb interaction constant:
$U=0$ (solid thin), $U=4\omega$ (solid thick), $U=8\omega$ (dotted), $U\rightarrow \infty$ (space dotted). 
The plots demonstrate that the number of FC steps in the interval $2\omega$ is doubled. Besides, when $U\ne0$  
the Coulomb blockade effects lead to suppression of the height of FC steps. For intermediate Coulomb
interaction, $U\simeq N_m\omega$ (see Eq.~(\ref{N_lambda}) in the text), the steps are strongly suppressed in height (solid thick curve), 
and for strong Coulomb interaction, $U \gg N_m\omega$, the FC steps for large bias voltages disappear (dotted curve and space dotted curve). 
The regime of sequential electron tunneling was considered,  $T=0.01\omega \gg\Gamma=0.0075 \omega$. 
The inset shows the dependence of differential conductance on bias voltage $G(V)/G_{\Gamma}$ for $U=4\omega$: 
the plot demonstrates that the dependence of peaks of differential conductance on
bias voltage is non-monotonic for strong Coulomb interaction
(manifestation of strong interplay of FC and Coulomb blockade
lifting). 
In the inset $T=0.1\omega,\, G_{\Gamma}=G_0\Gamma/(2\omega),\,G_0= e^2/2\pi$.  
b) The current-voltage characteristics when the Coulomb coupling constant is not a multiple
of $\omega$: $U=0.5\omega$ (solid thin), $U=2.5\omega$ (solid thick), $U\rightarrow \infty$ (dotted). 
Here $h=\omega$, other parameters are the same as in the Fig.~2a.  
The plots show that the doubling of steps takes place after the first stage of Coulomb blockade lifting. 
In the regime of Coulomb blockade ($U\rightarrow \infty$) the doubling does not appear.}
\end{figure}

\begin{figure}\centering
\includegraphics[scale=0.5]{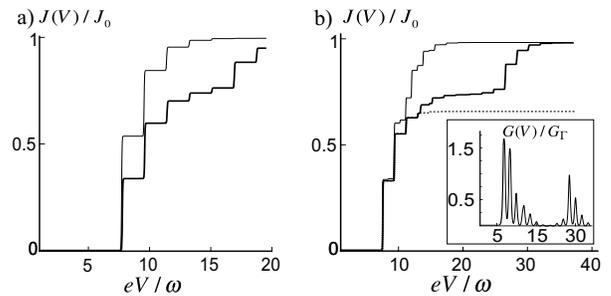}
\caption{ a)The current-voltage dependences of single-molecule transistor for strong external magnetic field, $h=7.25\omega$. 
Non-interacting limit $U=0$ (solid thin), and the case when  $U$ is a multiple of $\omega$, $U=5\omega$ 
(solid thick curve) are presented. 
In strong magnetic fields $h \gg \omega$ the doubling of the number of FC steps does not occur. 
Alternatively, in the case $U=0$ the steps 
are characterized by doubled heights.  b) The current-voltage dependences in strong external magnetic field, $h=7.25\omega$, 
for different Coulomb interaction constant: when $U/\omega$ is non-integer ( $U=2.5\omega$ (solid thin), $U=10.5\omega$ (solid thick)) and $U\rightarrow \infty$ (dotted). 
In the case when $U/\omega$ is non-integer the current-voltage characteristics  
show the doubling of FC steps after the first stage of Coulomb blockade lifting. In the regime of 
Coulomb blockade ($U\rightarrow \infty$) the steps are not doubled and they
disappear at high voltages. Besides, the plots show, that the suppression and disappearing of FC steps occurs also in
strong magnetic fields $h\gg \omega$ for large values of $U$ (solid thick curve).
All parameters are the same as in Fig.~2a. The inset shows the dependence of differential conductance on bias voltage $G(V)/G_{\Gamma}$ for $U=10.5\omega$. The appearance of two well separated conductance peaks (the energy interval is $e\delta V\simeq h$), superimposed by FC steps, is another manifestation of staged Coulomb blockade lifting. Here $h=7.25\omega$, all other parameters are the same as in the inset of Fig.~2a.}
\end{figure}

In our model when the level splitting $h\simeq \omega$ the number of FC steps per interval $2\omega$ is doubled compared to FC steps
in the model with spinless electrons (see e.~g.~\cite{Koch, Krive} for comparing).
The jumps of the current caused by the opening of inelastic 
channels "built" upon two elastic channels ($\varepsilon_{1,2}$) are shifted
by $h$ relatively to each other that results in steps doubling, in general case when $h$ is not a multiple of $\omega$. Indeed, for $U=0$ it is seen from Eqs. (\ref{K})-(\ref{Npm}) that at threshold voltage $eV=h$,
when elastic channels are open, $N_-=0,\;N_+=[h/\omega]$. If $h$ is not integer in $\omega$ units, the functions $K_-$ and $K_+$ are increased independently with the 
growth of $eV$. 
This is confirmed by numerical calculations which show that, analogously to the case $U=0$ (solid thin curve in Fig.~2a), steps doubling 
holds for $U\ne0$ (solid thick curve and dotted curve) 
and even when $U \rightarrow \infty$ (space-dotted curve, which practically
coincides with dotted curve at voltages $eV\lesssim 2U\approx 16\omega$). 

The doubling of the FC steps also happens after the first stage of 
Coulomb blockade lifting (Fig.~2b) in general case when $U/\omega$ is non-integer 
(and even if $h$ is a multiple of $\omega$). 
The reason is that at the threshold voltage $eV/2=\varepsilon_{\uparrow \downarrow}-\varepsilon_1=U-h/2$ for double occupation of the dot new set of current steps, associated with vibronic channels, arises. Nevertheless, at this bias voltage the double occupation of the dot
starting from the level $\varepsilon_2$ is not possible yet and the contribution of $\varepsilon_1$  corresponds to the value of $K_-$ with $N_-=[(U-h)/\omega]$ (we consider the case $U>h$ for definiteness). 
Let us consider that $h$ is a multiple of $\omega$. Then, if $U$ is not a multiple of $\omega$, the jump of current 
after the first stage of Coulomb blockade lifting does not coincide with the jump of the function $K_-$, i.e. the number of steps is doubled. 
It can be easily shown that the number of steps is doubled  in general case of non-integer $U/\omega$ for $U<h$. The doubling is clearly seen in the simulations in the Fig.~2b for $U<h$ (solid thin curve). 
When $U>h$ the number of steps are not doubled in
the first stage of
Coulomb blockade lifting (solid thick curve). For Coulomb
blockade regime the doubling due to $U$ is not observed ($U\rightarrow\infty$, dotted curve, which practically coincides with the solid thick curve at voltages $eV\lesssim 2(U-h/2)\approx 14\omega$). Note that in the simulations in the Fig.~2b $h$ is a multiple of 
$\omega$. 
In a general case the doubling of the number of FC steps is a special feature of our model. 

In the limit of strong fields $h\gg \omega$ the doubling does not occur (Fig.~3a) even if $h$ is not a multiple of $\omega$ (see Eq.~(\ref{Kh}), which is valid for non-interacting electrons). One more specific feature of current on voltage dependences in strong magnetic fields and $U=0$ is
the doubling of step heights (Fig.~3a, solid thin curve).
At $h \gg \omega$ the Coulomb correlations still affect the appearance of FC steps. The steps are not doubled when the value of $U/\omega$ is integer (solid thick curve in the Fig.~3a).  
The doubling of steps in the limit $h\gg \omega$ occurs when $U/\omega$ is non-integer (Fig.~3b, solid thin and solid thick curves).

We also obtain that for strong Coulomb interaction the current on voltage dependences has the regions characterized by the suppression of height or by full absence of the Franck-Condon steps. 
The plot for $U=0$ (solid thin curve in Fig.~2a) has the form of a sequence of doubled FC 
steps with monotonically decreasing heights. When Coulomb interaction 
$U \gtrsim \omega$ the $J(V)$ curve demonstrates two series of doubled steps with decreasing heights (Fig.~2a, solid thick).  
For strong Coulomb interaction 
$U \gg \omega$ (Fig.~2a, dotted curve) the region without steps appears. The suppression of vibration steps is caused by strong interplay of  FC and Coulomb blockade lifting. In order to describe the phenomena it is useful to introduce the number $N_m$ of inelastic channels (associated with one elastic 
channel) that give main contribution to the current. By using 
Eq.~(\ref{Kplusminus}) and Stirling formula one gets (when $\lambda \gg 1$)
\begin{equation}\label{N_lambda}
N_m \simeq [\text{e}\lambda^2]
\end{equation} 
(notice that here $\text{e}$ is the base of natural logarithm, the symbol $[...]$ denotes integer part). 
To reach FC blockade lifting for channels, corresponded to singly occupied dot the bias voltage has to exceed the value $eV\simeq N_m\omega$ (for definiteness a small external magnetic fields are considered, $h\ll\omega$). When the Coulomb interaction is strong enough, $U \gg N_m\omega$, the Coulomb blockade is not lifted at these voltages, thus the steps in the region $eV\in(0,N_m \omega)$ have the suppressed height comparing to the $J(V)$ curve at $U=0$. Till new channels are not available, new steps are fully absent (dotted curve in Fig.~2a), as well as in the case $U\rightarrow \infty$ for all high voltages (space-dotted curve in Fig.~2a).  The Coulomb blockade lifting due to increasing of the voltage gives rise to new series of FC steps, corresponded to channels of tunneling through doubly occupied dot (dotted curve in Fig.~2a). For intermediate Coulomb interaction, $U\simeq N_m\omega$, new series of steps occur befor FC blockade is fully lifted, then the curves represent two series of steps with decreasing height (solid thick curve in Fig.~2a). The current in the region without steps equals to the saturation current for Coulomb blockade regime,  $U\rightarrow \infty$, (i.e. the limit $V\rightarrow \infty$ of the Eq.~(\ref{K_infty})). It reads $J_s=2J_0 h^2/[3(h^2+\Gamma^2/3)]$, and this value is smaller than the maximal current for non-interacting electrons. Note, that definitions "strong" and "intermediate" are conditional and they depends on the value of $\lambda$ for each system. The inset shows the dependence of differential conductance on bias voltage, which depicts the suppression of the FC steps via the suppression of conductance peaks.

In
strong magnetic fields $h \gg \omega$ the suppression of FC steps is also present (see Fig.~3b, solid thick plot). For strong interaction
$U \gg \omega$ the lifting of Coulomb blockade occurs in two stages separated
by energy interval $e\delta V\simeq h \gg \omega$. Then $J(V)$ curve is roughly represented
by two regions of FC steps
separated by a large interval of bias voltages without any steps. 
The maximum current in this case
 is bigger than in the case $U \rightarrow \infty$, because the system is already in the second stage of Coulomb blockade lifting (compare solid thick and dotted curve in Fig.~3b). The inset in the Fig.~3b  shows the dependence of differential conductance on bias voltage. The appearance of two well separated conductance peaks (the energy interval is $e\delta V\simeq h$), superimposed by FC steps, is another manifestation of staged Coulomb blockade lifting.

\begin{figure}\centering 
\includegraphics[scale=0.5]{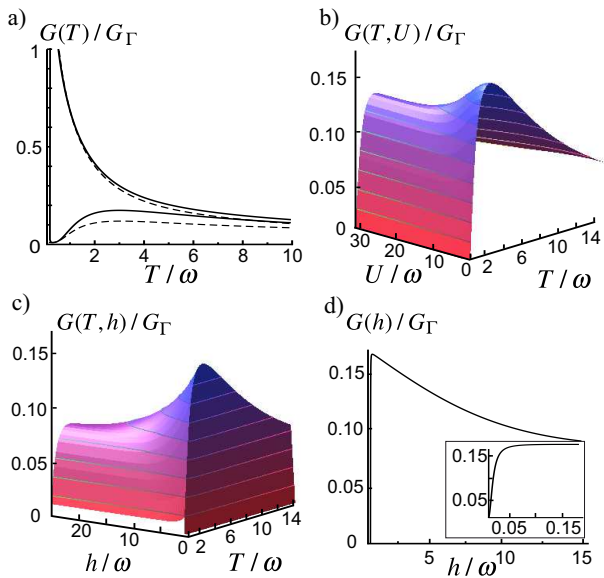}
\caption{a) The dependence of conductance $G(T)$ on temperature in absence of Coulomb interaction $U=0$. Two solid curves 
correspond to two values of electron-vibron interaction constant, $\lambda=1$ (solid top) and $\lambda=3$ (solid bottom). For strong
electron-vibron interaction, $\lambda\gtrsim 1.5$, the temperature
dependence of conductance is non-monotonic at intermediate temperatures
of the order of polaronic shift $\sim\lambda^2\omega$. Anomalous
temperature dependence is pronounced in FC blockade regime 
($\lambda\gtrsim 3$). Dashed lines present temperature dependence of
conductance for the same electron-vibron coupling but strong
Coulomb interaction $U=32\omega$. A slight suppression of anomalous
$T$-dependence is observed at temperatures $T\ll U$. The parameters chosen
in calculations are $h=0.05\omega$, $\Gamma=0.0075\omega$,
$G_{\Gamma}=G_0\Gamma/2\omega,\,G_0= e^2/2\pi$. 
b) Anomalous temperature growth of conductance as a function of Coulomb
interaction $U$. The increase of $U$ slightly suppresses conductance
maximum. Magnetic field is the same ($h=0.05\omega$) as it was used in calculations of plots a).
c) The dependence of anomalous growth of conductance on external
magnetic field in the absence of Coulomb interaction ($U=0$). 
With the increase of $h$ maximum conductance is slightly suppressed.
Both Coulomb and magnetic suppression of conductance is small and they do not change temperature
behaviour of conductance at temperatures when electron polaronic state in the
dot is destroyed by temperature.
d) The suppression of conductance at small magnetic fields
$h\simeq\Gamma$ for $T=3\omega$ and $\lambda=3$. 
The inset shows the curve in the range of very small magnetic fields 
$h\lesssim\Gamma$ when spin blockade is present.}

\end{figure}

Now we consider how the strong electron-vibron interaction influences the 
temperature dependence of maximal conductance. 
For a more simple system (tunneling of spin-unpolarized electrons through a vibrating single-level
QD), when Coulomb interaction is not crucial for electron transport, polaronic effects were
studied in~\cite{Monthioux,Krive}. Non-monotonic temperature
dependence of conductance was predicted~\cite{Krive} and observed~\cite{Monthioux}
in electron transport in carbon nanopeapod-based transistor. Here we
consider the influence of magnetic field and Coulomb interaction on
polaronic effects in spintronic transistor.

Two solid curves in Fig.~4a show the temperature dependence of maximal linear conductance in
the absence of Coulomb correlations ($U=0$). For strong electron-vibron interaction 
($\lambda \gtrsim 1.5$) $G(T)$ dependence is non-monotonic.
It is well-known, that for non-interacting electrons maximum (peak) conductance for 
sequential electron tunneling through a single-level QD scales
on temperature as $1/T$. Vibration effects, by forming polaronic state
of electron in the dot, strongly suppress conductance at low temperatures
$T \ll \omega$ (Franck-Condon blockade~\cite{Koch}). At high temperatures
$T \gg \lambda^2\omega$ polaronic state is destroyed and Franck-Condon blockade is
lifted. The lifting of polaronic blockade at intermediate temperatures results in
non-monotonic temperature dependence of conductance \cite{Krive}.

Small magnetic fields $h\ll\Gamma$ strongly suppress electron transport in
our spintronic transistor due to spin blockade. Therefore we will consider
magnetic fields $h\gtrsim\Gamma$ when spin blockade is lifted. 
The characteristic energy scale, when Franck-Condon blockade (we assume strong
electron-vibron coupling) is lifted, is determined by polaronic shift
$\lambda^2\omega$ (more precisely this energy scale is $N_m\omega$, see
Eq.~(\ref{N_lambda})). When magnetic field exceeds this energy scale the level $\varepsilon_1$ does not contribute to anomalous growth of conductance.
This means that the optimal range of magnetic fields to observe non-monotonic
temperature behaviour is $\Gamma\ll h\lesssim N_m\omega$. In this range of
magnetic fields the anomalous temperature behaviour of conductance, predicted
for spinless electrons, is weakly affected by magnetic field (peak conductance
slightly decreases with the increase of magnetic field). By the same reasons the Coulomb interaction has to be not too strong,
$U\lesssim N_m\omega$, in order to observe the anomaly when Coulomb blockade
is already lifted. 
Our qualitative considerations are supported by numerical
calculation of temperature dependence of conductance. The plots in Fig.~4a
demonstrate that temperature anomaly of conductance arises for sufficiently
strong electron-vibron interaction. Dependence of conductance maximum on
Coulomb interaction is shown in Fig.~4b. As it was expected, in the considered
range of Coulomb interaction strengths the maximum conductance weakly depends
on $U$. Analogously, magnetic fields in the range $\Gamma\ll h\lesssim N_m\omega$ 
slightly influence peak conductance (see Figs.~4c,d). Our numerical
analysis showed that non-monotonic temperature dependence of conductance,
which is the signature of strong electron-vibron interaction in the system,
can be observed in a wide range of external magnetic fields and even
for strong Coulomb interaction.
 
It is useful here to estimate the value of external magnetic field corresponded 
to maximal peak conductance in our model by using parameters of
fullerene-based transistor in the experiment~\cite{Park}. For  $\hbar \omega\approx5\,meV$ 
the optimal external magnetic field for observation of conductance 
temperature anomaly is of the order of $H\approx 1$T. 

\section{Conclusions}

In single-molecule transistors the vibration degrees of freedom can strongly
affect electron transport. 
In this paper we study the transport properties of spintronic transistor,
where the current of spin-polarized electrons between magnetic leads
is controlled by external magnetic field. Vibrations of a molecule
(quantum dot), which plays the role of base electrode in our transistor,
strongly modify current-voltage characteristics known for electron
transport in single-electron transistors. The interplay of Coulomb
blockade, Zeeman level splitting in external magnetic field and vibration
effects (Franck-Condon steps, polaronic blockade) results in non-trivial
behaviour of current on voltage dependences and temperature dependence of conductance.

Tunnel transport of spin-polarized electrons between magnetic leads with
opposite magnetization in ideal case of full spin-polarization is possible
only in the presence of non-parallel to magnetizations external magnetic
field, which induces electron spin-flips and lifts spin blockade.
The sensitivity of electron transport to weak magnetic fields makes our system
useful in spintronics. Vibration and Coulomb blockade effects diminish
electron current in single-molecule transistors and their influence on the
transport properties of single-electron spintronic transistor is not known.
The energy scale of Coulomb correlations and vibration-induced polaronic
effects can be of the same order of magnitude and these energies, as
a rule, are much larger than the Zeeman level splitting. In this case Coulomb and
Franck-Condon blockade is still present while spin blockade is totally
lifted. What are new features of charge transport in single-molecule
transistors in this regime?
We predict Coulomb blockade lifting (by bias voltage) occurred in stages and the doubling
of Franck-Condon steps. Both features are not known for spinless electrons. 
On contrary, anomalous temperature behaviour of conductance -- the signature of
strong electron-vibron interaction -- is weakly affected by Coulomb correlations
and is not changed in a wide region of magnetic fields.


 \textbf{\emph{Acknowledgement}}
 
This work is supported by the National Academy of Sciences of 
Ukraine (grant No. 4/19-N and Scientific Program 1.4.10.26.4)
 and partially by the Institute for Basic Science in Korea. 
 The authors thank O.A. Ilinskaya, A.V. Parafilo and R.I. Shekhter for 
 fruitful discussions. I.K. acknowledges
 the hospitality of PCS IBS in Daejeon (Korea).

\end{document}